\documentclass[sigconf]{acmart}

\usepackage{listings}
\preto\lstlisting{\def\@captype{table}}
\newenvironment{myfont}{\fontfamily{<pcr>}\selectfont}{\par}
\DeclareTextFontCommand{\textmyfont}{\myfont}

\AtBeginDocument{%
  \providecommand\BibTeX{{%
    \normalfont B\kern-0.5em{\scshape i\kern-0.25em b}\kern-0.8em\TeX}}}

\setcopyright{acmcopyright}
\copyrightyear{2020}
\acmYear{2020}
\acmDOI{
}

\acmConference[MLHat '20]{MLHat: The First International Workshop on Deployable Machine Learning for Security Defense}{August 24, 2020}{San Diego, CA}
\acmBooktitle{MLHat: The First International Workshop on Deployable Machine Learning for Security Defense, August 24, 2020, San Diego, CA}
\acmPrice{
}
\acmISBN{
}

\begin{document}

\title{MALOnt: An Ontology for Malware Threat Intelligence}

\author{Nidhi Rastogi}
\email{raston2@rpi.edu}
\orcid{0000-0002-2002-3213}
\affiliation{%
  \institution{Rensselaer Polytechnic Institute}
  \city{Troy}
  \state{New York}
}

\author{Sharmishtha Dutta}
\email{duttas@rpi.edu}
\orcid{0000-0002-2002-3213}
\affiliation{%
  \institution{Rensselaer Polytechnic Institute}
  \city{Troy}
  \state{New York}
}

\author{Mohammed J. Zaki}
\email{zaki@cs.rpi.edu}
\orcid{0000-0002-2002-3213}
\affiliation{%
  \institution{Rensselaer Polytechnic Institute}
  \city{Troy}
  \state{New York}
}

\author{Alex Gittens}
\email{gittea@rpi.edu}
\orcid{0000-0002-2002-3213}
\affiliation{%
  \institution{Rensselaer Polytechnic Institute}
  \city{Troy}
  \state{New York}
}

\author{Charu Aggarwal}
\email{charu@us.ibm.com}
\orcid{0000-0002-2002-3213}
\affiliation{%
  \institution{IBM T. J. Watson Research Center}
  \city{Yorktown Heights}
  \state{New York}
}
\renewcommand{\shortauthors}{Rastogi and Dutta, et al.}

\begin{abstract}
  Malware threat intelligence uncovers deep information about malware, threat actors, and their tactics, Indicators of Compromise (IoC), and vulnerabilities in different platforms from scattered threat sources. This collective information can guide decision making in cyber defense applications utilized by security operation centers (SoCs). In this paper, we introduce an open-source malware ontology - MALOnt that allows the structured extraction of information and knowledge graph generation, especially for threat intelligence. The knowledge graph that uses MALOnt is instantiated from a corpus comprising hundreds of annotated malware threat reports. The knowledge graph enables the analysis, detection, classification, and attribution of cyber threats caused by malware. We also demonstrate the annotation process using MALOnt on exemplar threat intelligence reports. A work in progress, this research is part of a larger effort towards auto-generation of knowledge graphs (KGs) for gathering malware threat intelligence from heterogeneous online resources.\end{abstract}



\maketitle

\section{Introduction}
Malware attacks impact every industry that is enabled by Internet technology --- approximately 7.2 billion malware attacks were reported worldwide in 2019\footnote{https://tinyurl.com/yxs8h6aw}. Such attacks cause loss, alteration, and misuse of sensitive data and compromise system integrity. Malware 
often have typical patterns corresponding to the type of industry they attack, groups of attackers behind similar attacks and means to pave their way into the target system, traces left behind after an attack has taken place, and so on. For preventing and detecting future attacks - both similar and disparate, the collection, integration, and analysis of malware threat intelligence is crucial.
\par
A malware ontology can support the construction of models that can detect and track attacks from their initial stages (such as identification of a vulnerability) through later stages (such as an exploit or data compromise). An ontology acts as a blueprint of a specific domain, containing key concepts (classes), their properties. Restrictions on classes are defined to limit the scope of the class, which is then inherited by the instances as well.

Therefore, it can facilitate the aggregation, representation, and sharing of threat information which would otherwise be challenging to reproduce, reuse, and analyze at a large scale. Both human and software agents can use an ontology to understand the structure of information that is stored in a document, a report, a blog, a tweet, or any other structured, semi-structured, or unstructured information source\cite{ontology101}.
\par
Specifically for malware threat, an ontology can provide a dictionary of attacks and related information that can help SOC analysts to dig deeper into their origination, attack goals, timeline, affiliated actors, vulnerabilities exploited for the attack, impact on industries as well as on humans. Such rich pieces of information can be aggregated following the ontology classes and data properties, which can significantly enhance current, future, and sometimes past analysis of online attacks, thereby curbing their propagation before a malware becomes hard to contain. In the absence of an ontology, security researchers and SoC analysts struggle to manage information from multiple sources and rely on ad-hoc mechanisms. There are existing attack and threat taxonomies that can be extended to a knowledge graph, however, they have their limitations that we cover in later sections. Linking common information between threat sources also becomes complex and therefore researchers are either compelled to look at attack instances in isolation or make do with available context.
\par
The main contribution of this paper is MALOnt - an open-source malware ontology\footnote{https://github.com/shoron-dutta/MALOnt} which underpins the collection of malware threat intelligence from disparate online sources. MALOnt contains concepts ranging from malware characteristics to attack and attacker details. For example, malware details may include family details, attack vectors (software or hardware vulnerabilities) deployed by an attacker, targeted operating system, impacted industries, history of attacks, and so on. MALOnt can be populated with specific instances and thus, be expanded to generate a knowledge graph (KG). A malware knowledge graph reasoner can infer new facts through deduction or induction and relies on machine learning and deep learning models to greatly expand the range and scale of fact generation. MALOnt will be used as a baseline framework for generating the KG, which is part of our ongoing research. MALOnt is implemented to supplement malware threat information extraction by following the steps below:

\begin{enumerate}
    \item Create MALOnt - an ontology for malware threat intelligence.
    \item Create a malware knowledge graph by integrating malware-related information with the ontology.
    \item Infer implicit intelligence from the knowledge graph using an OWL (Web Ontology Language) reasoner.
\end{enumerate}

These steps will be described in detail in Section \ref{implementation}.

\section{Background Concepts}
In this section, we cover key concepts that are used to create and instantiate MALOnt.

\subsection{Threat Reports}
 \begin{figure}[htp]
\centering
\frame{\includegraphics[width=0.45\textwidth, height=0.25\textwidth]{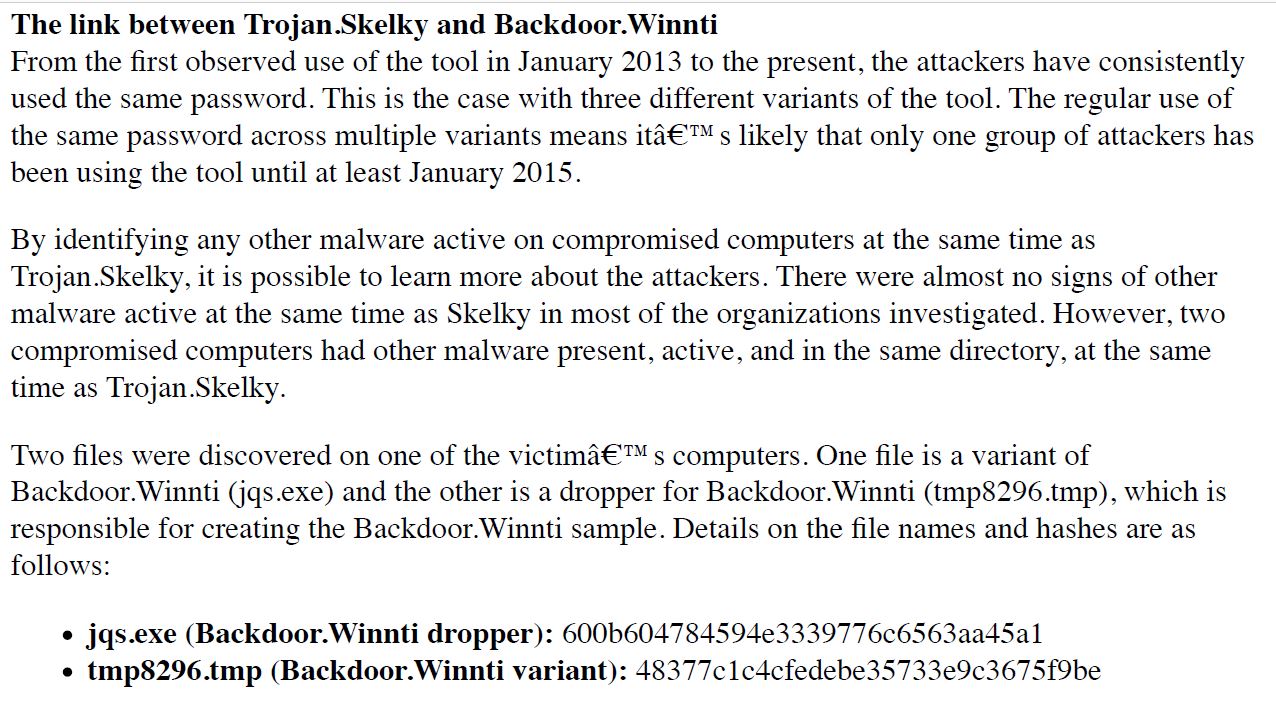}}
\caption{An excerpt of a threat report on Backdoor.Winnti malware family}
\label{fig:report_snapshot}
\end{figure}
When a malware attack occurs or when a software vulnerability is identified, a detailed account of these actions is captured by researchers and security analysts in threat reports. Corrective measures are eventually taken to prevent further propagation of the malware, and more evidence is added to the earlier documented accounts. Threat reports are technical in nature and cover the information related to malware (a collective name for viruses, trojan, ransomware, spyware) such as vulnerabilities exploited, operating system and applications impacted, modus operandi, group or cybercriminal responsible, hash of the malware, first sighting of the attack, determined IP addresses of attack server, and so on. Other security analysts and researchers utilize these reports as a reliable source to study and analyze malware samples, adversary techniques, exploited zero-day vulnerabilities, and much more. Figure \ref{fig:report_snapshot} shows a snippet from a sample threat report\footnote{\href{https://tinyurl.com/y9e7m5w7}{https://tinyurl.com/y9e7m5w7}} describing Backdoor.Win32 (Win64) malware family.
\subsection{Ontology}\label{subsection:ontology}
An ontology broadly describes concepts within a domain through classes and properties. These properties include property between defined classes and their attributes. An ontology is usually designed around a few main classes that cover the domain whose scope has been pre-defined. These classes may have sub-classes (more specific), super-classes (more general). The relationship between classes determines the type of interaction between them. Instances are individual instantiated examples of a class, which means each instances can have different values for data properties and be connected to other instances via object properties. Three classes - \text{Malware}, \textmyfont{Location}, and \textmyfont{AttackerGroup}, largely describe a malware's behavior. This can be vetted with the attack or vulnerability details captured in a few relevant threat reports. The \textmyfont{Malware} class has two sub-classes - \textmyfont{TrojanHorse} and \textmyfont{Dropper}. A property \textmyfont{hasTargetLocation} exists where the domain is determined by the \textmyfont{Malware} class and range by the \textmyfont{Location} class. Two properties exist \textit{from} \textmyfont{AttackerGroup} class \textit{to} \textmyfont{TrojanHorse} and \textmyfont{Dropper} classes titled \textmyfont{usesTrojan} and \textmyfont{usesDropper} respectively. A property of the \textmyfont{Dropper} class is titled \textmyfont{deliveredIn} that represents the mechanism of how the dropper is delivered to the target system. \par
To build an ontology, three main approaches are recommended while also engaging human expertise to build them:
\begin{enumerate}
    \item {\textit{Top-down}} - Classes are defined from the root of the class hierarchy by identifying the most general classes first\cite{ontologyEngineering, synthesisLectures, ontology101}. This approach is preferred when the goal of the ontology is to represent distinctive features of a domain \cite{ontologyEngineering}.
    \item {\textit{Bottom-up}} -  One starts with the leaves in the class hierarchy and builds higher levels of abstraction along the way \cite{ontologyEngineering, synthesisLectures, ontology101}. Relevant data sources can be used to identify concepts that are expressed in the dataset.
    \item{\textit{Middle-out}} - The most important classes are determined first followed by the rest of the class hierarchy. It is a combination of top-down and bottom-up approaches\cite{synthesisLectures, ontology101}.
\end{enumerate}

\begin{figure}[htp]
    \centering
   \frame{\includegraphics[width=.4\textwidth, height=.4\textwidth]{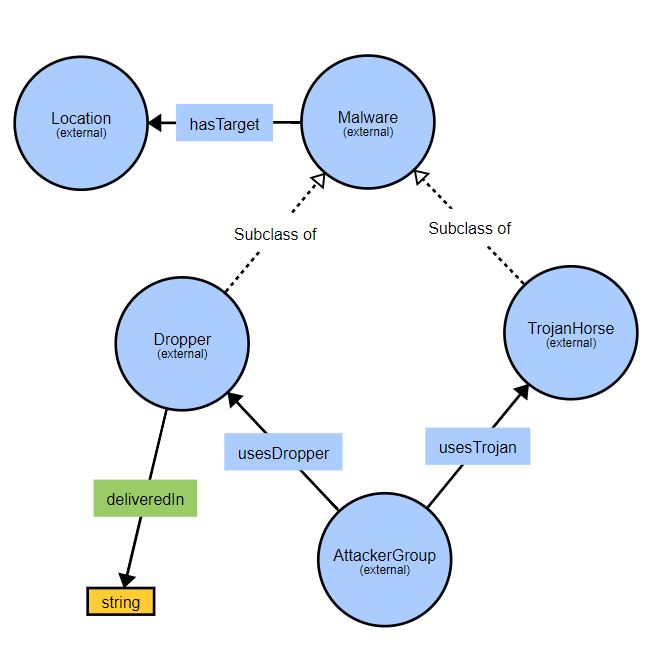}}
    \caption{Visual portrayal of a few top Classes in MALOnt using VOWL plugin in Protege.}
    \label{fig:snap}
\end{figure}

\par To elaborate, refer to Figure \ref{fig:snap}, where we describe a few concepts from MALOnt.
\subsection{Knowledge Graphs}

\begin{figure}[htp]
    \centering
   \frame{\includegraphics[width=.45\textwidth, height=.38\textwidth]{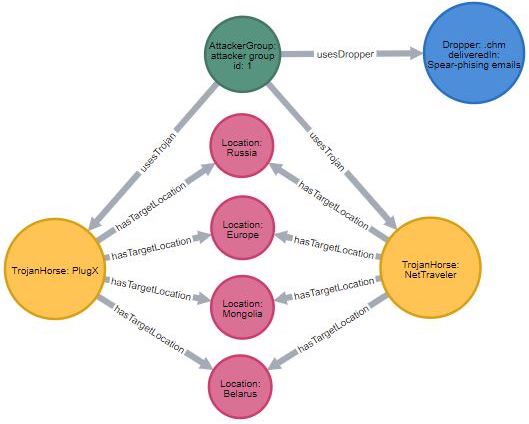}}
    \caption{Snippet of an exemplar malware knowledge graph using neo4j}
    \label{fig:kg_background}
\end{figure}

A knowledge graph is a machine-readable data repository containing a large amount of structured and unstructured data. This data is stored as triples $\langle\textit{Subject, Predicate, Object}\rangle$, where the predicate indicates the relationship between a subject and an object. Each entity or node in a knowledge graph has a unique identifier and may be connected via properties. A unique identifier allows capturing provenance information comprising references to threat sources of the triples. The graph structure can be exploited for efficient information extraction. Ontologies play a crucial role in building knowledge graphs (KGs). One way to build a KG is by adding individual instances to the ontology classes, and properties \cite{ontology101}. In addition to this, class and property instances outside of those defined by the ontology can be added to an existing KG, which allows for flexibility in KG generation. Consider the ontology explained in subsection \ref{subsection:ontology}. When a small portion of the ontology is populated with instances collected from threat reports, we get a small malware knowledge graph. Such as, the report titled ``Oops, they did it again: APT Targets Russia and Belarus with ZeroT and PlugX\footnote{https://tinyurl.com/yavqfb2y}'' contains information about an attacker group, which can be mapped to \textmyfont{AttackerGroup} class. The attacker uses trojans - \textit{PlugX} and \textit{NetTraveler}, to target infrastructures in Europe, Russia, Mongolia, Belarus, among others. This attacker group uses a dropper \textit{Microsoft Compiled HTML Help (.chm)} which is delivered via spear-phishing involving bespoke emails. This information is mapped to MALOnt classes (described in subsection \ref{subsection:ontology}), to generate a small part of the malware knowledge graph, as visualized in Figure \ref{fig:kg_background}. A KG is not just an instantiated data of an ontology. It is a web of properties between individual nodes (also called entities) and uses a reasoner to draw connections between entities that would otherwise not be understood.
\par
\begin{figure}[htp]
        \centering
        \frame{\includegraphics[width=0.45\textwidth]{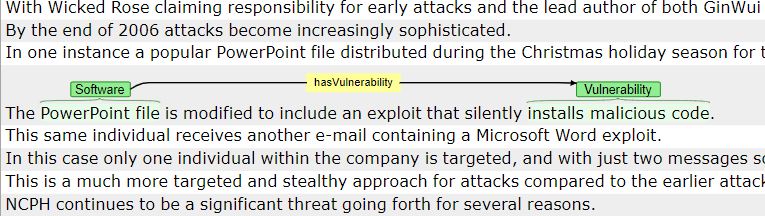}}
        \caption{An annotated threat report using brat \cite{brat}.}
        \label{fig:brat}
\end{figure}

Threat Reports are annotated (see Figure \ref{fig:brat}) using annotation tools such as Brat Rapid Annotation Tool\cite{brat}, and INCEpTION to create instances of classes defined in MALOnt. Here, the text segments ``PowerPoint file'' and ``installs malicious code'' are labeled as MALOnt classes titled \textmyfont{Software} and \textmyfont{Vulnerability} respectively. The arrow from \textmyfont{Software} to \textmyfont{Vulnerability} denotes the semantic relationship between these two classes \textmyfont{hasVulnerability}.

\section{Literature Review}
We corroborate the timeliness and necessity of MALOnt in this section by explaining the gaps in existing approaches and by comparing it with the state of the art standards, taxonomies, and ontologies.

\subsection{Existing Malware ontologies}
Swimmer\cite{swimmer} presented one of the first classifications of malware and their behavior using two classes: \textit{Malware} and \textit{MalwareCharacteristic}. Stucco\cite{stucco} expressed functionalities for capturing information on an attack although it lacked the means to store properties such as malware type(dropper, trojan, etc.) or attacker's location. Unified Cyber Ontology (UCO)\cite{uco} is based on STIX\cite{stix} and other cybersecurity standards and is mapped to vocabularies and sharing standards in the cybersecurity domain, as well as external sources such as DBPedia. To the best of our knowledge, the entire UCO is not publicly available and is broader in scope when compared to MALOnt. As normally practiced, some of the cybersecurity concepts for basic classes (e.g. \textmyfont{MalwareCharacteristic}\cite{swimmer}) were imported, however, MALOnt goes beyond describing just the malware attack behavior. It also captures the impacted industries, malware propagation mechanism, timeline, targeted system, prevention, and so on.
\par
One might argue against the need for a malware ontology since there exist quite a few standards and taxonomies, as well as ontologies in the cybersecurity domain that can be used to share malware threat intelligence in a structured way. We compare some of the most prominent ones with MALOnt here. Mitre’s Common Vulnerabilities and Exposures (CVE)\footnote{https://cve.mitre.org/} dictionary identifies publicly known security vulnerabilities in software packages. Common Attack Patterns Enumerations and Characteristics (CAPEC)\footnote{\label{note1}https://capec.mitre.org/} provides an enumeration of repeated techniques in cyber attacks. Mitre’s Adversarial Tactics, Techniques, and Common Knowledge (ATT\&CK)\footnote{https://attack.mitre.org/} provides a list of publicly known adversaries, their techniques, and post-compromise tactics to achieve their objectives. OpenIOC\footnote{https://www.fireeye.com/blog/threat-research/2013/10/openioc-basics.html} is a standard format for sharing IOCs. CVE\cite{cve}, CAPEC\cite{capec}, ATT\&CK, and IoCs provide static information of already discovered malware artifacts about malware attacks (among other information), which facilitates the representation and integration of collected information. STIX\cite{stix}, a knowledge representation standard, is expressed in XML which does not support reasoning or identifying properties between class instances. 
\par
The aforementioned standards cannot parse multitudes of threat advisory information and present it in a meaningful, human-readable, actionable format that can be used by AI models \cite{rastogi2018network} for prediction or analysis. It is our observation that the earlier work focuses on ontologies from specific threat vectors, such as malware. MALOnt differs from these largely because the domain is beyond threat Due to big data available in the cyber threat landscape, we strive to create information extraction techniques that can perform automated analysis, enable reasoning, enhance inference capabilities with minimal human intervention. This feature is currently lacking in available standards. Therefore, we propose the use of Web Ontology Language or OWL\footnote{https://www.w3.org/OWL/} as the language for malware threat knowledge representation and analysis.

\subsection{Knowledge Graphs for Malware}
Generating knowledge graphs for malware threat intelligence is an emerging research area. This is partly due to a limited background in KG and in adopting its concepts for security research. The paper closest to MALOnt and the proposed malware KG is \cite{CKG}\cite{RelExt}, where a pipeline to create knowledge graphs from after action reports (similar to threat reports) is proposed. Existing standards and vocabularies in cybersecurity were used in conjunction with the threat reports to prepare the training dataset for the cybersecurity KG. In contrast, a combination of vector spaces and a knowledge graph was proposed in \cite{VKG}. Vector embedding can be more efficient in searching similar nodes whereas knowledge graphs enable reasoning. These complementary strengths were used to build a pipeline comprising knowledge extraction, representation, and querying of data objects which performed better than its components.
\par
Aside from these, multitude domain-specific knowledge graphs have been built around an existing ontology or a generic knowledge base such as WikiData\cite{ceres, kbFromWikidata, wikidata}, and made openly accessible to the scientific community. While undoubtedly useful, they cater to generic concepts in the real world such as \textmyfont{Person, Organization, Location}. To the best of our knowledge, there exists no open-source knowledge graph in the malware threat intelligence domain that captures sufficient details to enable large scale automated malware threat analysis.

\section{Ontology Design \& Implementation} \label{implementation}
In this section, we describe the methodology and scope for defining MALOnt classes and properties, the requirement criteria, MALOnt's intended application, as well as example classes and properties.
\begin{figure}[htp]
        \centering
        \frame{\includegraphics[width=0.25\textwidth, height=0.3\textwidth]{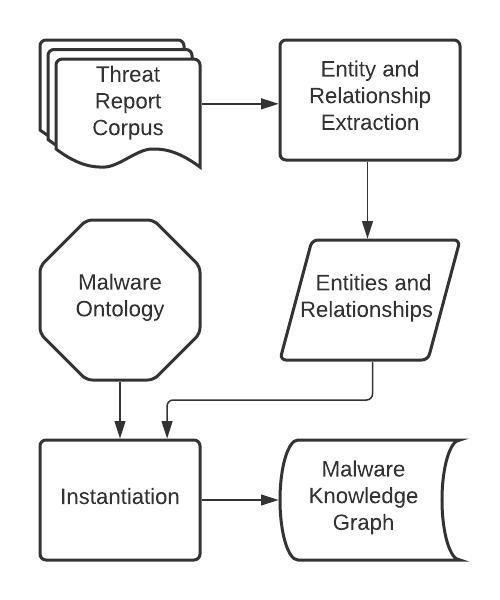}}
        \caption{Using a Malware Ontology to construct a Malware Knowledge Graph}
        \label{fig:architecture}
\end{figure}

\subsection{Purpose and intended use of MALOnt}
Malware threat reports are written in natural language and describe malware attacks in detail. Information retrieval from such data feeds can be unstructured in nature, which poses several challenges for information extraction.
\par An ontology can serve the purpose of mapping disparate data from multiple sources into a shared structure using a common vocabulary that would further facilitate the collection, aggregation, and analysis of that data\cite{BuildingAnOntologyOfCS}.
Therefore, we propose MALOnt - a malware ontology to encapsulate the concepts necessary to represent information regarding malware. The intended purpose of MALOnt is threefold: 
\begin{enumerate}
\item To capture semantic information from threat reports by assigning individual entities or instances to a pre-defined class in MALOnt and identifying properties where applicable. 
\item To use factual evidence present in the reports and infer new connections and properties between instances.
\item  To serve as a foundation for creating a malware KG by populating MALOnt with individual instances from threat reports.
\end{enumerate}

The steps required to achieve these goals are shown in a graphical format in Figure \ref{fig:architecture}. Once instantiated, MALOnt can extract information such as the indicators of compromise, adversary information, software vulnerabilities, attack tactics, and much more. 
It would also assist researchers and security analysts who gather malware intelligence from unstructured sources. Furthermore, software agents can utilize MALOnt to generate malware KGs.

\subsection{Competency Questions for MALOnt}\label{competency_questions}
Before creating an ontology, it's requirements \cite{ontology101} should be gathered, defined, and scoped by answering relevant competency questions. Having these competency questions act as the north star when identifying pertinent classes and properties for proper coverage of the domain. SPARQL queries can either be used to answer a question or a narrower version of a broad competency question by running them on the instantiated ontology. For MALOnt, the domain of the ontology is cybersecurity. In order to create the scope within the larger cybersecurity domain, over two dozen threat reports and existing ontology related sources (owl files, and research papers) were reviewed.  Key terms from the reports were identified and the hierarchy of existing ontologies was studied. This helped us vet out other ontologies, import relevant classes to MALOnt, and create the class hierarchy that adequately covers different aspects of malware threat intelligence. Below are the three competency questions that broadly cover the scope of malware threat intelligence:
\begin{enumerate}
\item Which malware characteristics adequately define malware threat landscape? (including methods, vulnerabilities, targets, and cybercriminals.)\label{CQ1}
\item What are the similar features for grouping adversaries, malware to help understand their behavior and predict the future course of action?\label{CQ2}
\item What is the impact of a given malware on an organization or industry? (financial, human life, intellectual property, reputation)\label{CQ3}
\end{enumerate}
\subsection{Creating MALOnt}


Designing and developing an ontology is an agile process. Three stages were continuously visited and reviewed based on core-competency questions - reviewing threat reports, identifying classes, hierarchy, data properties, and evaluating existing security ontologies. The middle-out approach in creating ontology also covers the first two stages. For MALOnt, many top-level classes were created with hierarchy and data-properties abstracted from threat reports. Instances were created for these classes by capturing individuals from threat reports. In the rest of this section, we use examples to describe the middle-out approach for building MALOnt.
\par 
The pre-defined upper-level classes such as \textit{Malware}, incorporate some of the most relevant details about a malware attack. \textmyfont{Host, Information, MalwareCharacteristics, Malware} are extracted from existing ontologies \cite{stucco, swimmer, insiderThreat}. The family of a specific malware is represented by \textit{MalwareFamily}. These two classes are joined by the property \textit{hasFamily} where \textit{Malware} class is its domain and \textit{MalwareFamily} class is the range. 

\par
Thereafter, we reviewed the competency questions defined in Section \ref{competency_questions} and identified classes such as \textit{Attacker}, \textit{Organizations}, and \textit{Indicator}. Over two dozen threat reports were manually reviewed to identify key concepts that needed to be included in the ontology as new classes or properties. For example, threat reports frequently provide valuable details about software vulnerabilities exploited by the malware, as well as specific release or version of the software product. In consequence, we included a \textit{Software} class with two properties \textit{hasReleaseYear} and \textit{hasVersion}. To connect the instance for class \textit{Software} to its vulnerability, \textit{hasVulnerability} property is introduced. 

\subsection{Exemplar Classes and Relations}
\begin{figure}[htp]
\centering
\frame{\includegraphics[width=8cm]{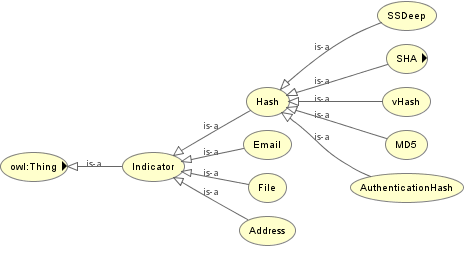}}
\caption{Indicator class in MALOnt using OWLviz plugin in Protege\cite{noy2001creating}.}
\label{fig:indicator_class}
\end{figure} 
In this section, we list out and describe some of the top-level classes and properties in MALOnt, which are essential to extract information from malware threat reports. 
\begin{itemize}
\item \textmyfont{Malware}: The general concept of malware, which is malicious software intended to violate the integrity, availability, or confidentiality of a computer system\cite{ctiModel}. It has four sub-classes.

\item \textmyfont{MalwareFamily}: A group of malware with common properties. Often threat reports describe the behavior of a malware family (see Figure \ref{fig:report_snapshot}) to help detect or prevent a novel malware belonging to that family.

\item \textmyfont{Attacker}: An adversary or a cybercriminal who can cause damage to a computer system by illegal methods. It is assumed that all attackers in this class are humans.

\item \textmyfont{AttackerGroup}: A group of cybercriminals who have homogeneous signatures of attack.

\item \textmyfont{ExploitTarget}: An entity (a person or an organization) that is the target of a malware attack.

\item \textmyfont{Indicator}: Distinguishable artifacts in a computer system that indicates malicious or suspicious behavior.

\item \textmyfont{Location}: Geographic location of a place.

\end{itemize}

The \textmyfont{Indicator} class construct can be seen in Figure \ref{fig:indicator_class}. Since an indicator of compromise (IoC) can be of different forms (file, email, hash, address), four sub-classes are created to define them. Furthermore, a malware hash signature is of different types, and the six sub-classes of the \textmyfont{Hash} cover them.
\par
Next, we describe a few properties that represent the semantics of the sentences and connect the instances in malware threat reports. The domain is used to define property characteristics whereas range enforces restrictions. Together they help maintain the integrity of the ontology, contain the possibility of inconsistencies, or erroneous conclusions by the automated reasoners.
\begin{itemize} 
\item \textit{hasVulnerability}: Bridges an exploit target or a software to its vulnerability.\\ \textbf{Domain}: \textit{ExploitTarget}, \textit{Software} \\ \textbf{Range}: \textit{Vulnerability}
\item \textit{hasAttachment}: Creates link from a malicious email to the attachment it contains\\ \textbf{Domain}: \textit{Email} \\ \textbf{Range}: \textit{File}
\item \textit{indicates}: Connects an indicator of compromise to its origin. It has an inverse relation titled \textit{indicatedBy}\\ \textbf{Domain}: \textit{Indicator}\\ \textbf{Range}: \textit{Malware}
\item \textit{usesDropper}: Connects a malware or an adversary to a frequently used measure  - a dropper.\\ \textbf{Domain}: \textit{Attacker}, \textit{Malware}, \textit{Campaign}\\ \textbf{Range}: \textit{Dropper}
\item \textit{hasFamily}: Maps a malware to its family. There is an inverse relation of this named \textit{hasMember}\\ \textbf{Domain}: \textit{Malware}\\ \textbf{Range}: \textit{MalwareFamily}
\item \textit{hasCharacteristics}: Maps a malware instance to its behavioral attributes \\ \textbf{Domain}: \textit{Malware}\\ \textbf{Range}: \textit{MalwareCharacteristics}
\end{itemize}

Class specific properties, called datatype properties, are shown in Figure \ref{fig:snap}. For example, the class \textmyfont{Software} has two properties- \textmyfont{hasVersion} and \textmyfont{hasReleaseYear}.


\section{Evaluation}
In this section, we evaluate MALOnt by running SPARQL queries on the ontology. These SPARQL queries answer to specific use cases of the competency questions, explained in Section \ref{competency_questions}. This method of evaluation is referred to as goal modeling \cite{fernandes2011using} and is considered a very effective evaluation technique to test the adaptability and consistency of an ontology \cite{ontology_evaluation_survey}. If the SPARQL queries are able to extract instances as a response, it signifies that the competency questions have succeeded in covering the defined goal of the ontology.

\begin{enumerate}
\item \textbf{Retrieving threat information related to malware characteristics.} Competency question \ref{CQ1} can have a specific use case, where MALOnt  is queried to extract attributes of different malware campaigns. MALOnt's property \textmyfont{targets} is selected from \textmyfont{Campaign} to \textmyfont{Organization} to get a list of all malware campaigns, their respective target organizations as well as persons (see SPARQL query in Listing\ref{listing:query_object_property}).
\begin{lstlisting}[basicstyle=\footnotesize,
label={listing:query_object_property}, caption=SPARQL Query for Competency Question \ref{CQ1}
]
SELECT DISTINCT ?instance ?p ?o 
WHERE {
    ?instance a ?x .
    ?instance ?p ?o .
    ?p a owl:ObjectProperty .
    ?x a owl:Class .
    ?x rdfs:label 
        "Campaign"^^xsd:string .
    ?p rdfs:label "targets" 
    }
\end{lstlisting}

\item \textbf{Retrieving similar features for grouping concepts}. Competency question no.\ref{CQ2} can take various forms. Here, we show a SPARQL query that leverages the inverse properties of classes in MALOnt, to find all malware families whose member malware have left a specific IoC footprint. In Figure \ref{fig:malware_family}, each instance of \textmyfont{Malware} class in MALOnt is mapped to an instance that belongs to class \textmyfont{MalwareFamily} using property titled \textmyfont{hasFamily}. Alternatively, there is an inverse relation of \textmyfont{hasFamily} called \textmyfont{hasMember}.  
\textmyfont{Malware} instances can also be mapped to \textmyfont{Indicator} instances using \textmyfont{indicatedBy} property. The inverse of this property is \textmyfont{indicates}.

\begin{figure}[htp]
    \centering
    \includegraphics[width=.45\textwidth, height=.3\textwidth]{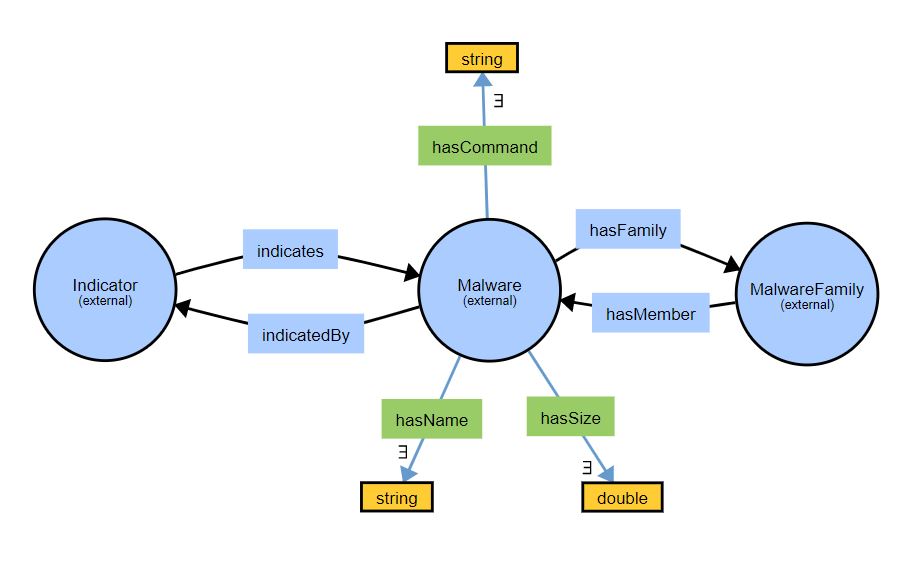}
    \caption{Inverse properties in MALOnt using VOWL plugin in Protege.}
    \label{fig:malware_family}
\end{figure}

\par Once an IoC (defined by class \textmyfont{Indicator}) of a given malware is extracted from a threat report, the property is identified, which traces the malware back to the malware family. More insights about a malware family can be gathered through such chain properties between the three MALOnt classes. A SPARQL query is executed for detecting instances of \textit{MalwareFamily} that have any member malware indicated by an \textit{Indicator} class. The SPARQL query traverses two kinds of triple structures, $\langle \textit{MalwareFamily, hasMember, Malware}\rangle$ and $\langle \textit{Malware, indicatedBy, Indicator}\rangle$ in order to find the response, see listing \ref{listing:query_chain}:

\begin{lstlisting}[basicstyle=\footnotesize,
label={listing:query_chain}, caption=SPARQL Query for Competency Question \ref{CQ2}
]
SELECT DISTINCT ?malware_family ?p
                    ?malware ?q ?t 
WHERE {
    ?malware_family ?p ?o .
    
    ?malware_family a ?x.
    ?x a owl:Class.
    ?x rdfs:label "MalwareFamily"
        ^^xsd:string.
    ?p a owl:ObjectProperty.
    ?p rdfs:label "hasMember".
    
    ?malware ?q ?t .
    
    ?malware a ?z .
    ?z a owl:Class.
    ?q a owl:ObjectProperty .
    ?q rdfs:label "indicatedBy"
        ^^xsd:string .
    ?t a owl:NamedIndividual .
    
    ?t rdfs:label "indicator_value"
}
\end{lstlisting}

\item \textbf{Retrieving information on affected person or organization}. For competency question no. \ref{CQ3}, one can extract information about affected systems, organizations, or person(s). The SPARQL query in listing \ref{listing:query_CQ3} retrieves a list of target objects, and accessed information from those objects by a specific attacker group. This query can retrieve all information of a particular \textmyfont{AttackerGroup} entity using properties where \textmyfont{AttackerGroup} is the domain.

\begin{lstlisting}[basicstyle=\footnotesize,
label={listing:query_CQ3}, caption=SPARQL Query for Competency Question \ref{CQ3}]
SELECT DISTINCT ?instance ?p ?o ?q
WHERE {
	?instance ?p ?o .
	?instance a ?x .
	?instance rdfs:label 
	    "AttackerGroup1"^^xsd:string .
	?p a owl:ObjectProperty .
    	?x a owl:Class .
   	?p rdfs:label ?q .
	?o a ?object .
	?object a owl:Class .
	
 }
\end{lstlisting}

\end{enumerate}

\section{Application of MALOnt}

In this section, we demonstrate the process of creating a part of the KG (due to space constraints) by instantiating MALOnt with over a dozen threat reports. We also explain how the reasoner can be used to retrieve information by capturing it from multiple threat reports by executing SPARQL queries on the exemplar KG.

\subsection{Annotating Threat Reports}
MALOnt has been instantiated with open-source threat reports\footnote{\href{https://github.com/CyberMonitor/APT_CyberCriminal_Campagin_Collections/}{https://github.com/CyberMonitor/APT\_CyberCriminal\_Campagin\_Collections/}} that were published between the years 2006 to 2020. Many of these reports have been published by reputed organizations working within the cybersecurity domain. These reports provide a range of coverage on malware threats prominent at the time of publication. A few other reports focus on homogeneous attributes of various attacks caused by malware. For example, a report \footnote{\href{https://www.nerc.com/pa/rrm/bpsa/Alerts\%20DL/2011\%20Alerts/A-2011-02-18-01\%20Night\%20Dragon\%20Attachment\%201.pdf}{https://www.nerc.com/pa/rrm/bpsa/Alerts\%20DL/2011\%20Alerts/A-2011-02-18-01\%20Night\%20Dragon\%20Attachment\%201.pdf}} published in 2011 covers details on a set of operations called as Night Dragon. Another report\footnote{\href{https://github.com/CyberMonitor/APT_CyberCriminal_Campagin_Collections/blob/master/2013/energy-at-risk.pdf}{https://github.com/CyberMonitor/APT\_CyberCriminal\_Campagin\_Collections/\\blob/master/2013/energy-at-risk.pdf}}  published in 2013, focusses on Night Dragon, Stuxnet, and Shamoon. Annotating different kinds of reports in the corpus allows deeper and wider range of details on a particular malware attack.
\begin{figure}[htp]
        \centering
       \includegraphics[width=.33\textwidth, height=.33\textwidth]{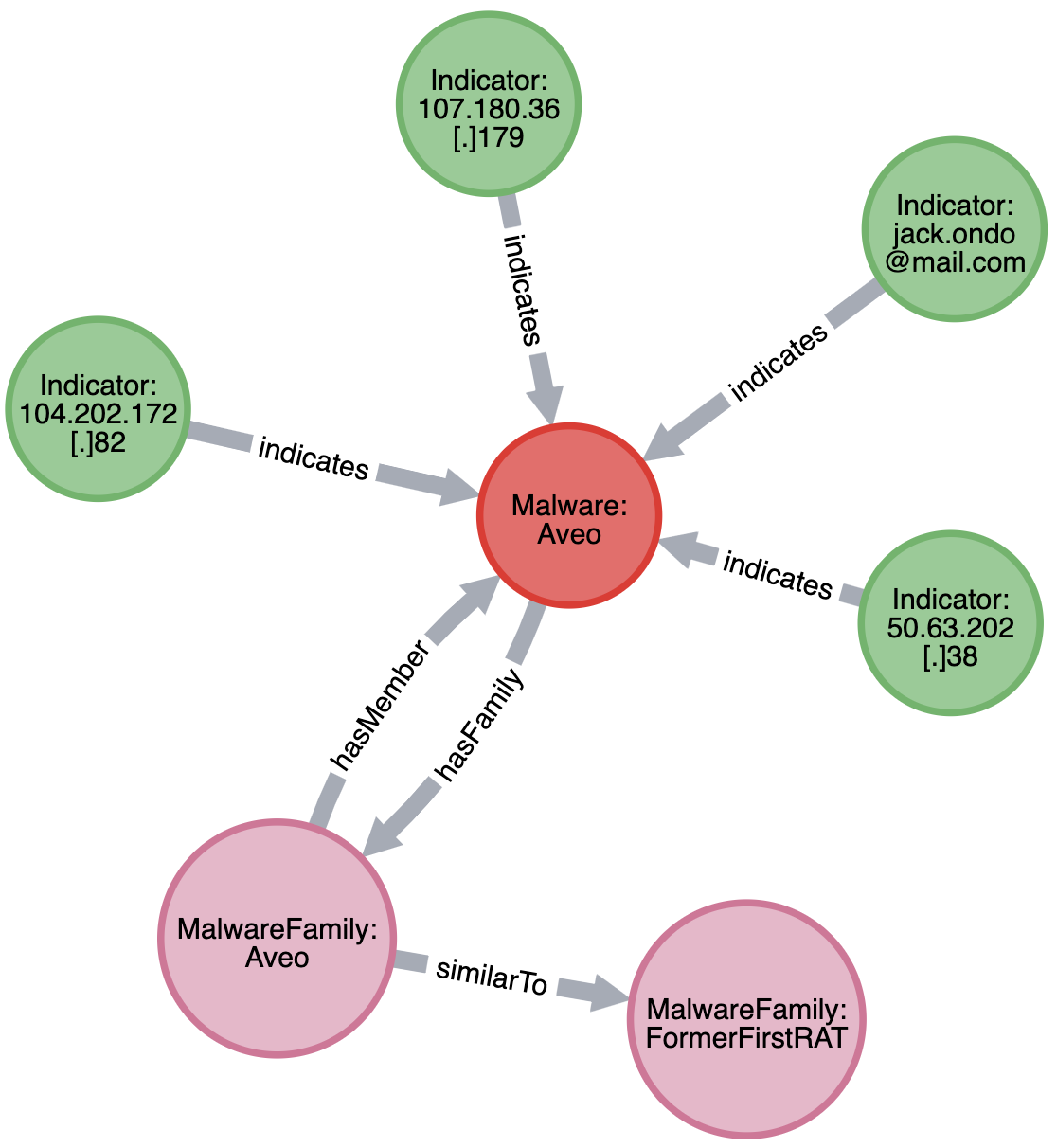}
        \caption{Exemplar knowledge graph based on MALOnt in neo4j}
        \label{fig:kg}
    \end{figure}
\par
Threat reports were manually annotated by the authors and reviewed by a security expert. The annotated values from the threat reports were used to instantiate various concepts of MALOnt. In this step, values are assigned to instances of MALOnt classes and properties. In Figure \ref{fig:kg} the snippet of the malware KG depicts modeling of threat data collected from reports using classes and properties from the MALOnt ontology.

\section{Conclusion and Future Work}
In this paper, we propose MALOnt - an ontology for malware threat intelligence by defines 68 classes, 31 properties, and 13 properties for representing malware attacks. While this is work in progress, we used the middle-out approach for creating ontologies to review the top classes as well as classes. As future work, MALOnt will further formalize the implicit assumptions of the malware threat domain in order to build a sustainable knowledge graph. For this, annotated malware threat reports will be reviewed and instantiated for MALOnt classes.
\begin{acks}
The authors would like to thank Shruthi Chari and Dr. Oshani Seneviratne for evaluating MALOnt and for ensuring that best practices are followed for ontology generation; and Destin Yee for putting together the ontology and knowledge graph figures, and instantiating threat intelligence reports.
\end{acks}
\bibliographystyle{ACM-Reference-Format}
\bibliography{sample-sigconf}
\appendix
\end{document}